\documentclass[aps,twocolumn,amsmath,amssymb,showpacs,prl]{revtex4}
\usepackage{graphicx}

\newcommand{\etal}{{\it et al.}}

\begin{document}

\title{Non-dispersive Fermi arcs and absence of charge ordering in the pseudogap phase of Bi$_2$Sr$_2$CaCu$_2$O$_{8+\delta}$}
\author{
        U. Chatterjee,$^{1,2}$
        M. Shi,$^{1,3}$
        A. Kaminski,$^4$
        A. Kanigel,$^1$
        H. M. Fretwell,$^4$
        K. Terashima,$^5$ T. Takahashi,$^5$
        S. Rosenkranz,$^2$
        Z. Z. Li,$^6$
        H. Raffy,$^6$
        A. Santander-Syro,$^6$
        K. Kadowaki,$^7$
        M. R. Norman,$^2$
        M. Randeria,$^8$
        and        
       J. C. Campuzano$^{1,2}$
        }
\affiliation{
     (1) Department of Physics, University of Illinois at Chicago, Chicago, IL 60607\\
     (2) Materials Science Division, Argonne National Laboratory, Argonne, IL 60439 \\
     (3) Swiss Light Source, Paul Scherrer Institut, CH-5232 Villigen, Switzerland\\
     (4) Ames Laboratory and Department of Physics and Astronomy,\\
	         Iowa State University, Ames, IA 50011\\
     (5) Department of Physics, Tohoku University, 980-8578 Sendai, Japan\\
     (6) Laboratorie de Physique des Solides, Universite Paris-Sud, 91405 Orsay Cedex, France\\ 
     (7) Institute of Materials Science, University of Tsukuba, Ibaraki 305-3573, Japan\\
     (8) Department of Physics, Ohio State University, Columbus, OH  43210\\
         }
\date{\today}
\begin{abstract}
The autocorrelation of angle resolved photoemission data from the high temperature superconductor Bi$_2$Sr$_2$CaCu$_2$O$_{8+\delta}$ shows distinct peaks in momentum space which disperse with binding energy in the superconducting state, but not in the pseudogap phase. Although it is tempting to attribute a  non-dispersive behavior in momentum space to some ordering phenomenon, a de-construction of the autocorrelation reveals that the non-dispersive peaks arise not from ordering, but rather from the tips of the Fermi arcs, which themselves do not change with binding energy.
\end{abstract}

\maketitle
A central issue in condensed matter physics is the competition between different order parameters. This competition is most evident in the transition metal oxides, which show a rich variety of insulating, conducting, magnetic and superconducting phases. In the case of the high temperature cuprate superconductors, the situation is particularly interesting, with a much debated pseudogap phase lying between the insulating state at zero doping and the d-wave superconducting state. Recent Fourier transform scanning tunneling spectroscopy (FT-STS) experiments on Bi$_2$Sr$_2$CaCu$_2$O$_{8+\delta}$ have been interpreted \cite{VERSHININ} in terms of local charge ordering in the pseudogap phase. Here, using an autocorrelation analysis of angle resolved photoemission spectroscopy (ARPES) data, we demonstrate that the vectors observed in FT-STS originate from the high density of states  at the tips of the dispersionless Fermi arcs, and not from charge ordering.

The samples employed in this work are single crystals and thin film samples of Bi$_2$Sr$_2$CaCu$_2$O$_{8+\delta}$ grown using the floating zone method and RF sputtering technique, respectively. The samples were mounted with the Cu-O bond direction parallel to the photon polarization, and cleaved in situ at pressures below our measurement limit of  2x10$^{-11}$ Torr.  The incident photons were not aligned along a mirror plane to avoid strong selection rules. Measurements were carried out at the Synchrotron Radiation Center in Madison, Wisconsin, on the U1 undulator beamline, at 2 meV intervals, using a Scienta SES 2002 electron analyzer. A momentum resolution of 0.0036 $\AA^{-1}$ was used for the thin films, and 0.01 $\AA^{-1}$ for the single crystals, both with a photon energy of 22 eV, with data covering up to four times the irreducible zone in reciprocal space.  Both values of the momentum resolution, at least along the detector cut, are comparable to or better than that obtained by FT-STS, which for the largest field of view studied is about 0.01 $\AA^{-1}$. It should be noted that the resolution values quoted are along the momentum multiplexing direction of the detector, but these detectors have worse momentum resolution along the perpendicular direction ($\sim 0.02 \AA^{-1}$).  Finally, the spectra were normalized at high binding energies to eliminate the effect of the dipole matrix element on the intensities.

When producing the 2D autocorrelation maps, we took into account the fact that the data are taken over a finite range of momenta, since the number of data points in the autocorrelation sum contributing to a given vector q decreases with increasing q. We deal with this in two different ways. The first method is to normalize the autocorrelation function by the number of points used in the sum. The second method is to impose periodic boundary conditions, although we do not display these here. Another aspect of the data from the single crystals is the presence of photoelectron diffraction images (superlattice signal) that gives rise to ghost features in both the intensity maps and their autocorrelations. For these samples, we solve this problem by using only data in the Y quadrant, where the superlattice contribution is well separated from the main state, and reconstruct the whole Brillouin zone by reflections, which require interpolation of the data to a uniform grid. The large background emission in ARPES reduces the contrast in the autocorrelation maps. We deal with this by simply zeroing out the intensity once it falls below a given threshold value, which we take to be 20\% for the superconducting phase and 38\% for the pseudogap phase, but we emphasize that this procedure does not change the results.

\begin{figure}
\includegraphics[width=3.4in]{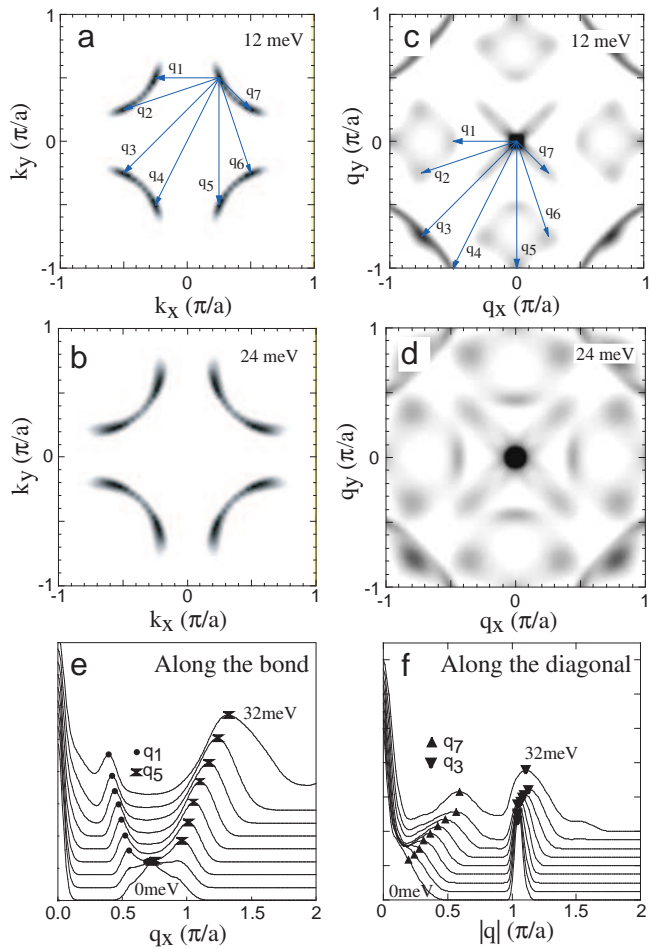}
\caption{Spectral intensities and their autocorrelations in the superconducting state ($T=40$ K) for an optimally doped single crystal sample. Intensity at a binding energy of (a) 12 meV and (b) 24 meV in the first zone.  (c)-(d) Autocorrelation of the data in (a)-(b), with vectors in (c) indicating local maxima in the autocorrelation which correspond to equivalent vectors shown in (a). The gray scales in these plots are in arbitrary units. (e) Autocorrelation of the data as a function of binding energy along the $(q,0)$ direction. The lowest curve is at zero energy, with the binding energy increasing by 4 meV for each subsequent curve (the curves are displaced vertically for clarity). (f) Same as (e), but along the $(q,q)$ direction.  For both directions, strongly dispersing peaks are seen. }
\label{fig1.eps}
\end{figure}

 In Fig.~1a and b, we show the ARPES intensity maps,  $I(\textbf{k},\omega)$, as a function of the momentum \textbf{k} in the first zone, for two different choices of the binding energy  $\omega$. These data are obtained for an optimally doped single crystal sample ($T_c = 90$ K) in the superconducting state ($T = 40$ K). The characteristic ÒbananaÓ shapes of the constant energy contours due to the d-wave energy gap can be clearly seen, with the size of the bananas increasing with increasing binding energy. In the right panels c and d, we plot the autocorrelation $C(\textbf q,\omega)=\sum_{\textbf k} A(\textbf{k+q},\omega) A(\textbf{k},\omega)$ of the ARPES intensities, i.e., the product of intensities at two different momenta at fixed energy, separated by a momentum transfer $\textbf{q}$, summed over the Brillouin zone. 
 
As we shall show in the remainder of this paper, the autocorrelation, which is effectively the momentum-resolved joint density of states, permits us to gain important new insights. $C(\textbf q,\omega)$ clearly exhibits discrete spots in q-space, very similar to peaks seen in FT-STS experiments \cite{HOFFMAN}. We note that the ARPES autocorrelation does not require any theoretical modeling for its interpretation, and the observed $\textbf{q}$-space pattern can be directly interpreted from the measured ARPES intensity  $I(\textbf{k},\omega)$. By restricting the autocorrelations over specific regions of the intensity maps (as we shall show), we can unambiguously determine the origin of those spots.
 
 This analysis reveals that the spots in the autocorrelations, Fig.~1c (and d), originate from the corresponding $\textbf{q}$ vectors spanning the high intensity points of Fig.~1a (and b respectively). In other words, the autocorrelation highlights regions of high density of states as can be seen by matching the spots in Fig.~1c, e.g., with the arrows spanning the highest intensity areas in Fig.~1a. We also see that the characteristic $\textbf{q}$-vectors disperse as a function of the binding energy  $\omega$. This dispersion is shown in detail in Fig.~1e for the vectors $\textbf{q}_1$ and $\textbf{q}_5$ along the bond direction, and in Fig.~1f for the vectors $\textbf{q}_3$ and $\textbf{q}_7$ along the zone diagonal. These autocorrelation spots always follow the end of the bananas at each  $\omega$. Remarkably, these $\textbf{q}$-vectors are in quantitative agreement with those obtained by the Davis group \cite{HOFFMAN,MCELROY1}, and therefore confirm that the ``octet" model \cite{WANGLEE} used to explain the FT-STS results does indeed describe the regions of highest density of states. It is surprising that FT-STS shows the same $\textbf{q}$-vectors, given that the formula used to theoretically model the FT-STS data, $[Im \sum_{\textbf k}  \hat G(\textbf{k+q},\omega) \hat T(\textbf{k+q},\textbf{k}, \omega) \hat G(\textbf{k},\omega)] _{11}$   (where $\hat{G}$  are Nambu Greens functions), which describes the scattering of electrons from the state {\textbf k} to the state {\textbf {k+q} by the scattering matrix  $\hat T$, is not mathematically equivalent to the simple autocorrelation   used in our analysis. 
 
We now turn to the much less understood pseudogap phase. In Fig.~2 we show data similar to that shown in Fig.~1, but now in the pseudogap phase  at $T=90$ K of a thin film sample with a $T_c=76$ K. The most significant feature of the intensity plots in Figs.~2a and b is that the ``Fermi arcs" \cite{ARCS}, the locus of gapless excitations in the pseudogap phase, do not significantly change with binding energy, in sharp contrast to the d-wave superconducting bananas \cite{MCELROY1}. This unexpected behavior is evident in a comparison of the arc lengths in the ARPES intensity maps in Figs.~2a,b. The autocorrelation of these data along the bond direction are shown in Fig.~2c and exhibit clearly identifiable peaks near  $\textbf{q}^*=(0.4\pi/a,0)$. Further, as shown in Fig.~2c these autocorrelation peaks show very little dispersion, in marked contrast to the superconducting state results of Fig.~1. This is the same wavevector observed by FT-STS \cite{VERSHININ} in the pseudogap phase. 
 
\begin{figure}
\includegraphics[width=3.4in]{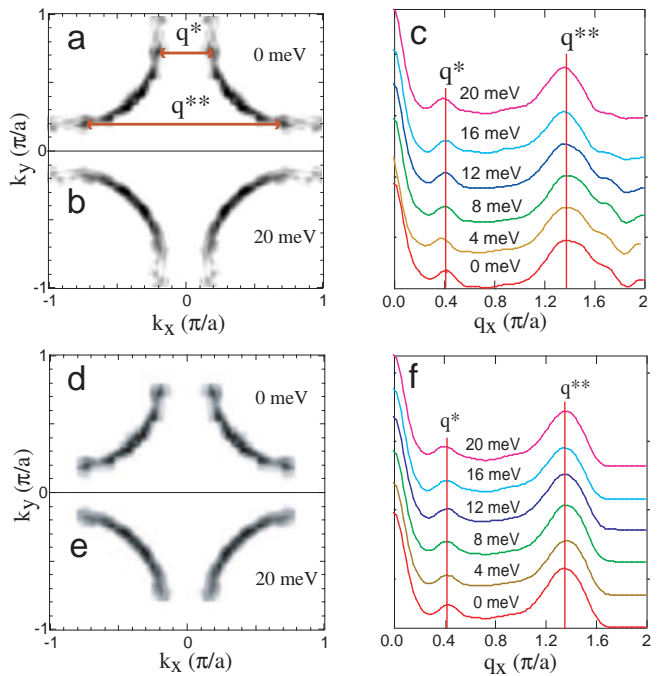}
\caption{Fermi arcs in the pseudogap state and their autocorrelations. Spectral intensity in the top half of the first zone at (a) 0 meV and (b) 20 meV for an underdoped thin film sample ($T_c = 76$K) at $T = 90$ K. One can see that the length of the arcs remains unchanged over the energy scale of the pseudogap. (c) Autocorrelation as a function of binding energy along the (q,0) direction. Note the dispersionless nature of the peaks labeled $\textbf{q}^*$ and $\textbf{q}^{**}$. (d) Spectral intensity as in (a), but with regions beyond the Fermi arcs cut off from the data. (e) Spectral intensity as in (b), but with the data cut off as in (d). (f) Autocorrelation of the data but with data beyond the Fermi arcs cut off, as shown for two energies in (d) and (e). Note that the results are unchanged from (c) for the  $\textbf{q}^*$ and $\textbf{q}^{**}$ peaks, indicating that the peaks in the autocorrelation originate from the ends of the arcs, and not from the pseudogapped sections of the Fermi surface.
}
\label{fig2.eps}
\end{figure}

We now use our intensity data to understand the origin of the dispersionless autocorrelation peaks seen in the pseudogap phase. We find that these peaks are located at vectors connecting the ends of the Fermi arcs \cite{ARCS}, where there is a sudden onset of the pseudogap. This onset is sufficiently steep so the Fermi arcs have nearly constant length for all binding energies smaller than the pseudogap value. This gives rise to the dispersionless nature of the autocorrelation vectors. We arrive at this conclusion by truncating the intensity plots to include only the Fermi arcs, as shown for 0 and 20 meV in Fig.~2d and e. We find that the autocorrelation of these data, shown in Fig.~2f, yields the same result as in Fig.~2c. Therefore, we conclude that $\textbf{q}^*$ connects the ends of the Fermi arcs, rather than the parallel sections near the zone faces, whose faint traces can be seen in Fig.~2a and b. This observation also explains why if a $\textbf{q}^*$ spanning the shorter distance between arc tips is present, $\textbf{q}^{**}$, spanning the longer distance between arc tips along the bond direction, must also be present, as seen in our autocorrelation plot along the bond direction in Fig.~2c. The latter $\textbf{q}^{**}$ does not appear in the data of Vershinin \etal \cite{VERSHININ}, but has been seen in later measurements (A. Yazdani, private communication). We also find additional peaks along the diagonal direction, as shown in the autocorrelation surface plot in Fig.~3d. These originate from vectors analogous to $\textbf{q}_3$ and $\textbf{q}_4$ in the superconducting state, which span the ends of the arcs along or near the diagonal directions. We have no explanation at present why these peaks are absent in the FT-STS data, though we note that in the superconducting FT-STS data, the intensity of these peaks drops sharply with reduced doping \cite{MCELROY2}. 

While the octet models and variants have been proposed to explain the FT-STS observations in terms of scattering of electrons by impurities in the superconducting  state \cite{HOFFMAN,WANGLEE,CAPRIOTI,MARKIEWICZ}, these models do not describe the pseudogap data \cite{FRANZ}. On the other hand, an autocorrelation analysis of ARPES data does not require a model for its interpretation, which leads us to our most significant finding: all features -- in both the superconducting and pseudogap cases -- have a common origin. They are associated with the $\textbf{k}$-dependent high density of states, either at the ends of the constant energy contours in the superconducting state (the ``bananas"), as first suggested by Hoffman \etal \cite{HOFFMAN}, or at the tips of the Fermi arcs in the pseudogap phase, as we show here. This is in spite of the dramatic differences between the autocorrelation maps in the superconducting state, shown in Fig.~3b and in the pseudogap phase, shown in Fig.~3d. 

Comparing our findings to those of FT-STS is useful, in that it directly addresses the question of charge ordering. In principle, collective effects can appear in the FT-STS maps \cite{KIVELSON,POLKOVNIKOV}. But in ARPES, these effects would show up as so-called ``shadow" bands, that is, images of the ``bananas" or ``arcs" translated by the presumed ordering wavevector (the $\textbf{q}^*$ of Fig.~2c), as indicated by the dashed curves in Fig.~3c. But clearly, we find no evidence for these shadows in Fig.~3c at the wave vectors in question in the FT-STS and ARPES autocorrelation maps. The difference between our conclusions and those of Ref.~\cite{MISRA} is the result of their assumption that the Fermi arc length increases with energy, which it does not in the actual ARPES data.
 
\begin{figure}
\includegraphics[width=3.4in]{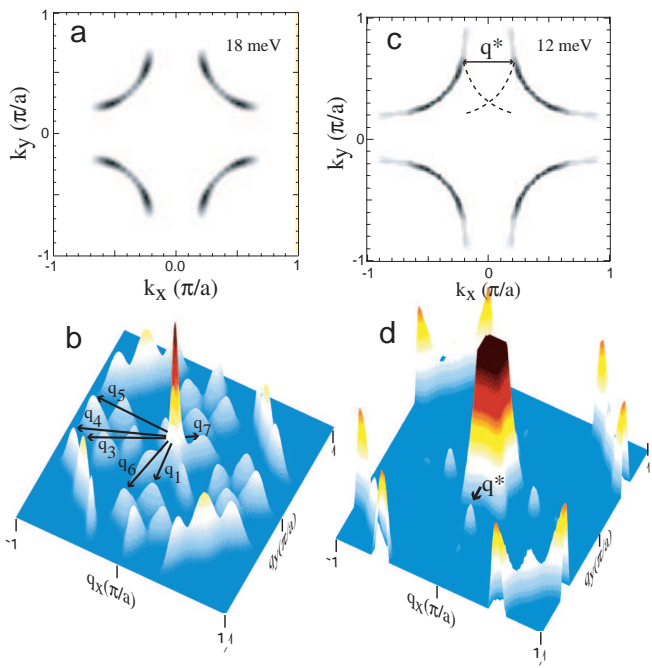}
\caption{Comparison of the resulting autocorrelations of similar  intensity maps in the superconducting and pseudogap phases of a single crystal sample ($T_c = 90$ K). (a) Intensity map in the superconducting state at $T = 40$ K at 18 meV. (b) Surface plot of the autocorrelation function from (a). (c) Comparison of an equal length Fermi arc in the pseudogap state ($T = 140$ K) at 12 meV to the superconducting banana in (a) obtained from the same sample. The dashed lines in (c) represent the arcs displaced by the wavevector $\textbf{q}^*$ identified in Fig.~2a, which are not present in the intensity maps, but would be  if $\textbf{q}^*$ was due to  charge ordering. (d) Surface plot of the autocorrelation function from (c). Note the significant differences between the two autocorrelation maps, (b) and (d), despite the similarities of the intensity maps, (a) and (c).
}
\label{fig3}
\end{figure}

In summary, while the dispersionless $\textbf{q}^*$ spots in the pseudogap phase have previously been interpreted as evidence for charge ordering \cite{VERSHININ,MCELROY1}, we show that this dispersionless behavior is simply a joint density of states effect. The joint spectral density has peaks at $\textbf{q}$-vectors which connect points of large density of states, namely at the ends of the constant energy contours (``bananas") in the superconducting state, and at the tips of the Fermi arcs in the pseudogap state. The ARPES autocorrelation peaks are in quantitative agreement with the FT-STS results both above and below $T_c$. The peaks are dispersionless in the pseudogap phase because the length of the Fermi arc itself changes very little with binding energy. Therefore we argue that the peaks in the FT-STS data must also arise from density of states effects, and not charge ordering. It is an interesting open question whether or not the low temperature FT-STS data on highly underdoped samples of a different material, Ca$_{2-x}$Na$_x$CuO$_2$Cl$_2$ \cite{HANAGURI}, can also be understood within this picture.

We acknowledge useful discussions with J.C. Davis and A. Yazdani. This work was supported by NSF DMR 0305253, the U.S. DOE, Office of Science, under Contract No. W-31-109-ENG-38, and the MEXT of Japan. The Synchrotron Radiation Center is supported by NSF  DMR-0084402.  Ames Laboratory is operated for the U. S. DOE by Iowa State University under contract No. W-7405-Eng-82.

\end{document}